\documentclass{article}  
\usepackage{natbib}
\usepackage{amssymb}
\usepackage{amsmath}

\begin{document}

\title{A Dutch Book against Sleeping Beauties Who Are Evidential Decision
  Theorists\thanks{This paper appears in {\em Synthese}, Volume 192, Issue
    9, pp.~2887-2899, October 2015.  The final publication is available at
    Springer via http://dx.doi.org/10.1007/s11229-015-0691-7}} 
\author{Vincent Conitzer\\Duke University}
\date{}

\maketitle

\begin{abstract}
  In the context of the Sleeping Beauty problem, it has been argued that
  so-called ``halfers'' can avoid Dutch book arguments by adopting
  evidential decision theory.  I introduce a Dutch book for a variant of
  the Sleeping Beauty problem and argue that evidential decision theorists
  fall prey to it, whether they are halfers or thirders.  The argument
  crucially requires that an action can provide evidence for what the agent
  would do not only at other decision points where she has exactly the same
  information, but also at decision points where she has different
  but  ``symmetric'' information.
\end{abstract}

\newpage
\section{Introduction}

The Sleeping Beauty problem (reviewed below) has attracted much attention
because it relates to a variety of unresolved philosophical
problems.\footnote{\cite{Titelbaum13:Ten} gives a useful overview of these
  connections.}  It is, in the first place, a puzzle about beliefs.  The
question of what Beauty should believe upon awakening divides philosophers
and others into multiple camps, mainly ``halfers'' and ``thirders'' (though
finer distinctions can be made).  But decision theory has also been pulled
into the debate.  This is because one natural strategy for adjudicating
between the halfer and thirder positions is to evaluate the effects of the
candidate beliefs on Beauty's decisions; if one position results in clearly
irrational decisions, this would appear to settle the matter.  As it turns
out, the success of such arguments appears to hinge on which version of
decision theory -- causal or evidential -- Beauty adopts.  A natural
reaction is to feel disappointment at the fact that a clean resolution to
the original question appears elusive.  But it is also possible to see
opportunity: perhaps variants of the Sleeping Beauty problem actually allow
us to adjudicate between causal and evidential decision theory instead.
This is the aim of this paper.  Of course, this may also help us resolve
the original question, insofar as the arguments against halfing or thirding
become decisive once the debate between causal and evidential decision
theory has been settled in this context.

Let us recall the standard variant of the Sleeping Beauty
problem~\citep{Elga00:Self}.  Beauty is put to sleep on Sunday, and a fair
coin is tossed.  If it lands Heads, she will be briefly awoken on Monday
only.  If it lands Tails, she will be briefly awoken on Monday, and then
again on Tuesday.  The table in Figure~\ref{fi:SB} summarizes this.
\begin{figure}
\begin{center}
\begin{tabular}{|r|c|c|}\hline
& Heads ($1/2$) & Tails ($1/2$)\\ \hline
Monday & awake & awake\\ \hline
Tuesday & asleep & awake\\ \hline
\end{tabular}
\end{center}
\caption{Coin tosses and corresponding awakenings, as well as
their probabilities, in the Sleeping Beauty problem.}
\label{fi:SB}
\end{figure}
Whenever she is awoken, she does not remember any previous awakenings, nor
does anything in the room indicate to her what day it is.  Beauty knows all
this throughout.  Now let us put ourselves in the shoes of Beauty when she
has just been awoken in the experiment.  What should be her credence that
the coin came up Heads?  Halfers believe that the answer is $1/2$; after
all, the fact that she has been awoken should not tell her anything about
the possible (uncentered) world that she is in, because the awakening event
is consistent with each of the two possible worlds.  In contrast, thirders
believe that the answer is $1/3$; after all, only $1/3$ of the possible
awakening events occur in a Heads world, and this would be borne out when
repeating the experiment many times.  Who is right?

One decision-theoretic approach to answering this question is to attempt to
construct a {\em Dutch book} to which one of the two positions is
vulnerable.  A Dutch book is a set of bets that the agent in question would
all accept individually, but that together ensure that the agent incurs a
strict loss overall.  In a {\em diachronic} Dutch book, the bets are
offered at different times; all of the Dutch books discussed in this paper
are diachronic.  \cite{Hitchcock04:Beauty} describes a Dutch book argument
against the halfer position.  In this Dutch book, the bookie first --
before Beauty goes to sleep -- sells her a bet that costs $15$ and pays out
$30$ if the coin lands Tails.\footnote{To minimize clutter, I will not
  specify currency units such as dollars or euros.}  Then, each time that
Beauty awakens, he sells her a bet that costs $10$ and pays out $20$ if the
coin has landed Heads.  The idea is that if Beauty is a halfer, then she
will always be willing to accept these bets.  (In fact, she will be
indifferent between accepting them and not accepting them; it is
straightforward to slightly modify the payoffs so that she will strictly
prefer to accept them.)  Now, if the coin lands Heads, she will buy bet 1
once and bet 2 once, at a total cost of $25$, and bet two will pay out $20$
-- so she will run a net loss of $5$.  On the other hand, if the coin lands
Tails, she will buy bet 1 once and bet 2 {\em twice}, at a total cost of
$35$, and bet 1 will pay out $30$ -- so again she will run a net loss of
$5$.  Thus the Dutch book succeeds.  The table in Figure~\ref{fi:Hitchcock}
summarizes Hitchcock's Dutch book.

\begin{figure}
\begin{center}
\begin{tabular}{|r|c|c|}\hline
& Heads ($1/2$) & Tails ($1/2$)\\ \hline
Sunday & bet 1: -15 & bet 1: 15\\ \hline
Monday & bet 2: 10 & bet 2: -10\\ \hline
Tuesday & no bet & bet 2: -10\\ \hline
total gain from accepting all bets & -5 & -5\\ \hline
\end{tabular}
\end{center}
\caption{The table shows which bet is offered when, as well as the net gain
  from accepting the bet in the corresponding possible world, for
Hitchcock's Dutch book.}
\label{fi:Hitchcock}
\end{figure}

However,~\cite{Draper08:Diachronic} point out that this Dutch book does not
succeed against a halfer that accepts evidential decision theory, because,
as~\cite{Arntzenius02:Reflections} pointed out earlier, such an agent would
calculate the expected utility of her options differently in the Sleeping
Beauty problem.  Specifically, consider the situation where Beauty accepts
evidential decision theory, has just been awoken, and is now calculating
the expected value of accepting bet 2.  For this calculation, by EDT, she
assumes that, if the coin has come up Tails, she also has accepted or will
accept bet 2 on the other day.  Thus, in this case, accepting bet 2 (on
both days) comes at a cost of $20$, and pays out nothing, for a net loss of
$20$; whereas in the Heads case, accepting bet 2 (once) comes at a cost of
$10$ and pays out $20$, for a net gain of only $10$.  So even if her
credence in Heads is as high as $1/2$, she will not accept bet 2, and the
Dutch book fails.\footnote{Draper and Pust do propose a modified Dutch book
  that involves telling Beauty that it is Monday and then offering a bet;
  this Dutch book works against some halfers, whether they are causal or
  evidential decision theorists, but not against so-called
  ``double-halfers'' who hold that the correct credence remains $1/2$ even
  after Beauty is told it is Monday.}

\cite{Briggs10:Putting} goes further and presents a proof that thirders who
accept causal decision theory and halfers who accept evidential decision
theory are both immune to Dutch books, and not just in the original
Sleeping Beauty problem but also in variants thereof.  The intuition for
why these two combined positions would be equivalent in this sense is the
following.  When an evidential decision theorist calculates her utility
conditional on accepting a bet (or not accepting it) in a possible world,
she assumes that she does the same upon every other awakening, so the
effect is multiplied by the number of times she awakens in that possible
world.  On the other hand, for a thirder, the probability of a possible
world is multiplied by the number of times she awakens in that possible
world.  This ends up having the same effect on the relevant expected
utility calculations.\footnote{The idea that thirding goes hand in hand
  with CDT, and halfing with EDT, also finds support elsewhere, for example
  in the context of the absentminded driver
  problem~\citep{Schwarz14:Lost}.}

Briggs' proof implicitly assumes that Beauty, upon an awakening, always has
the same information available to her.  This rules out variants such as
``Technicolor Beauty''~\citep{Titelbaum08:The} where Beauty, upon
awakening, sees a piece of paper whose color is not always the same.  The
Technicolor Beauty variant illustrates that irrelevant additional
information, such as the paper's color -- which is uncorrelated with the
variable of interest -- can change the halfer's credence for the variable
of interest (at least under a standard interpretation of halfing -- but see
Footnote~\ref{fo:halfing_interpretations}).  Indeed, Briggs does discuss
Technicolor Beauty and concludes that this instability of the halfer's
credence is a mark against halfing, but this is only after she has
completed her discussion of Dutch books.  What has not yet been
appreciated, to my knowledge, is the following: if Beauty is an evidential
decision theorist, then in variants such as Technicolor Beauty where she
does not always have the same information available to her upon waking, she
is vulnerable to Dutch books, regardless of whether she is a halfer or a
thirder.  This is what I will demonstrate in what follows.  I first
introduce a variant of the Sleeping Beauty problem.

\section{The White-Black-Grey (WBG) Sleeping Beauty Variant}

Beauty will be awoken twice, once on Monday and once on Tuesday.  As usual,
when she is awoken on Tuesday, she has no memory of the previous awakening.
The only information available to her upon awakening (besides the
information that was available to her at the start of the experiment) is
the color of the room in which she is awoken.  Two distinct fair coins are
tossed to determine the color of the room in which she is awoken on each of
the two days.  Coin 1 has a white side and a black side; this coin will be
used to determine the color of the room in which she is awoken on Monday.
Coin 2 has a grey side and a side with the word ``opposite'' written on it;
this coin will be used to determine the color of the room in which she is
awoken on Tuesday, where Opposite indicates that the color of the room
should be the opposite of what it was on Monday.  That is, if the first
coin comes up White and the second coin comes up Opposite, then she will be
awoken in the black room on Tuesday, and vice versa.  The table in
Figure~\ref{fi:WBG} shows the resulting possibilities for the sequence of
rooms in which Beauty is awoken, and their probabilities.\\

\begin{figure}
\begin{center}
\begin{tabular}{|r|c|c|c|c|}\hline
& WG ($1/4$) & WO ($1/4$) & BO ($1/4$) & BG ($1/4$)\\ \hline
Monday & white & white & black & black\\ \hline
Tuesday & grey & black & white & grey\\ \hline
\end{tabular}
\end{center}
\caption{Sequences of coin tosses and corresponding room colors, as well as
their probabilities, in the WBG Sleeping Beauty variant.}
\label{fi:WBG}
\end{figure}

I take it to be clear (though see
Footnote~\ref{fo:halfing_interpretations}) that, upon awakening in the
white room, Beauty should place $1/3$ credence in each of the centered
worlds WG/Monday, WO/Monday, and BO/Tuesday.  Specifically, I take it that
both halfers and thirders will agree on this, because in each of the three
possible worlds that have not been ruled out, there is only one awakening
(out of two) in the white room.  Similarly, upon awakening in the black
room, Beauty should place $1/3$ credence in each of BG/Monday, BO/Monday,
and WO/Tuesday.  Note the total symmetry between black and white in this
example, which will be essential to my argument.

In fact, the WBG variant is arguably isomorphic to Technicolor
Beauty~\citep{Titelbaum08:The}.  In Technicolor Beauty, the setup is the
same as in the original variant of the Sleeping Beauty problem, except an
additional coin is tossed.  If it comes up one way, Beauty sees a red piece
of paper on Monday and (if she wakes up on Tuesday at all) a blue piece of
paper on Tuesday; otherwise, the order in which she sees the colored pieces
of paper is reversed.  This Technicolor coin corresponds to the White /
Black coin in our WBG variant, and the original Sleeping Beauty coin
corresponds to the Grey / Opposite coin.  Staying asleep on Tuesday
corresponds to the grey room in the WBG variant.  I will stick with the WBG
variant here, in part because I believe it brings out the symmetry slightly
better, but also, and perhaps more importantly, because it ensures that
Beauty always awakens twice.  This latter property makes it difficult to
imagine that any credences other than those of $1/3$, $1/3$, $1/3$
described above could reasonably be considered correct.  If one agrees that
the corresponding credences are also correct for both halfers and thirders
in Technicolor Beauty,\footnote{\label{fo:halfing_interpretations}A
  standard interpretation of how the halfer assigns credences in general
  (e.g.,~\cite{Halpern06:Sleeping,Meacham08:Sleeping,Briggs10:Putting})
  would indeed, in the Technicolor Beauty variant, assign credences of
  $1/3$, $1/3$, $1/3$ in the possible worlds HR, TR, TB upon seeing a red
  piece of paper (where the first letter indicates the outcome of the
  original coin toss and the second letter the color seen on Monday), as
  these are the possible worlds that are not ruled out by the evidence.
  This, surprisingly, results in a credence of $1/3$ in Heads for the
  halfer.  However,~\cite{Pittard15:When} has objected to this conclusion
  and suggested that another interpretation of halfing should be found that
  keeps the credence in Heads at $1/2$ in Technicolor Beauty.  One
  interpretation of halfing that would achieve this is to treat Beauty's
  awakening as selected uniformly at random from her awakenings in the
  experiment in the actual world.  Under this interpretation, we would have
  $P(\mbox{see red} | \mbox{HR}) = 1$ (because HR has only one awakening)
  but $P(\mbox{see red} | \mbox{TR}) = P(\mbox{see red} | \mbox{TB}) = 1/2$
  (because in each of TR and TB, only one of two awakenings results in
  seeing red).  Hence, by Bayes' rule,
  $$P(\mbox{HR} | \mbox{see red}) = \frac{P(\mbox{see red} | \mbox{HR})
    P(\mbox{HR})}{ P(\mbox{see red} | \mbox{HR}) P(\mbox{HR}) + P(\mbox{see
      red} | \mbox{TR}) P(\mbox{TR}) + P(\mbox{see red} | \mbox{TB})
    P(\mbox{TB})} $$
  $$ = \frac{1 \cdot (1/4)}{1 \cdot (1/4) + (1/2) \cdot (1/4) + (1/2) \cdot
    (1/4)} = 1/2$$
  as desired by Pittard. In contrast, if we apply this interpretation of
  halfing to the WBG variant, we still obtain
$$P(\mbox{WG} | \mbox{see white}) = \frac{P(\mbox{see white}|
  \mbox{WG})P(\mbox{WG})}{P(\mbox{see white}| \mbox{WG})P(\mbox{WG}) +
  P(\mbox{see white}| \mbox{WO})P(\mbox{WO}) + P(\mbox{see white}|
  \mbox{BO})P(\mbox{BO})} $$ $$ = \frac{(1/2) \cdot (1/4)}{(1/2) \cdot
  (1/4) + (1/2) \cdot (1/4) + (1/2) \cdot (1/4)} = 1/3$$ as desired for the
Dutch book presented in this paper.  The key difference from Technicolor
Beauty is that $P(\mbox{see white}| \mbox{WG}) = 1/2$ because there is also
an awakening in the grey room.  Of course, the standard interpretation of
halfing also results in credences of $1/3$, $1/3$, $1/3$ in the WBG
variant.  The point of discussing this other interpretation of halfing here
is not to argue for it, but rather merely to show that while
interpretations of halfing may disagree about the correct credences in
Technicolor Beauty, it is hard to imagine that they would disagree in the
WBG variant.  One possible approach to finding an interpretation that
disagrees is to take Bostrom's approach of classifying awakenings into
``reference classes''~\citep{Bostrom02:Anthropic,Bostrom07:Sleeping} and
argue that the awakenings in the white and black rooms belong to the same
reference class, but not those in the grey room, thereby eliminating the
grey room from the picture in the calculation above.  However, it seems
hard to justify this classification without reference to the particular
details of the bets offered, and it seems difficult to swallow that
credences should depend on these details.}  then all that follows can also
be put in terms of that variant.  We are now ready to introduce the Dutch
book.

\section{A Dutch Book for the WBG Sleeping Beauty Variant}

Beauty will be offered the following bets.
\begin{itemize}
\item {\em Bet 1.}  This bet will be offered once, right before the
  experiment.  It costs $20$ and pays out $42$ if coin 2 comes up Grey.
\item {\em Bet 2.}  This bet will be offered once each time that Beauty
  awakens in the white or the black room, but never in the grey room.  Thus, it
  will be offered once overall if coin 2 comes up Grey, but twice overall
  if coin 2 comes up Opposite.  It costs $24$ and pays out $33$ if coin 2 comes
  up Opposite.
\end{itemize}
It should be noted that these bets are legitimate in the sense that the
bookie is not exploiting any information that Beauty does not have
available to her.  I will revisit this point in
Subsection~\ref{su:legitimate} below.

First, let us verify that accepting all these bets is sure to result in a
loss.  If coin 2 comes up Grey, then bet 2 is offered only once, so that
Beauty pays $20 + 24 = 44$, and receives a payout of $42$ from bet 1 -- so
she runs a loss of $2$.  On the other hand, if coin 2 comes up Opposite,
then bet 2 is offered twice, so that Beauty pays $20 + 2 \cdot 24 = 68$,
and receives a payout of $2 \cdot 33 = 66$ from the two iterations of bet 2
-- so again she runs a loss of 2.  The table in Figure~\ref{fi:Dutch_EDT}
summarizes the Dutch book.

\begin{figure}
\begin{center}
\begin{tabular}{|r|c|c|c|c|}\hline
& WG ($1/4$) & WO ($1/4$) & BO ($1/4$) & BG ($1/4$)\\ \hline
Sunday & bet 1: 22 & bet 1: -20 & bet 1: -20 & bet 1: 22\\ \hline
Monday & bet 2: -24 & bet 2: 9 & bet 2: 9 & bet 2: -24\\ \hline
Tuesday & no bet & bet 2: 9 & bet 2: 9 & no bet\\ \hline
total gain from accepting all bets & -2 & -2 & -2 & -2\\ \hline
\end{tabular}
\end{center}
\caption{The table shows which bet is offered when, as well as the net gain
  from accepting the bet in the corresponding possible world, for
the Dutch book presented in this paper.}
\label{fi:Dutch_EDT}
\end{figure}

But who will actually accept these bets?  Both causal and evidential
decision theorists will accept bet 1, because before the experiment there
is a 50\% chance that coin 2 comes up Grey, so that the expected payout
from bet 1 is $21$, which is greater than $20$.  Will a causal decision
theorist accept bet 2?  No: given that the room is (say) white, she
believes that there is a probability of $2/3$ that coin 2 has come up
Opposite, so the expected payout of the bet is $(2/3) \cdot 33 = 22$, which
is less than the cost of the bet, $24$.  So the causal decision theorist is
not vulnerable to this Dutch book.

All that remains to show is that the evidential decision theorist will
accept bet 2 whenever it is offered to her.  Here, then, is the crux of the
argument.  Suppose the room is white.  Then, accepting the bet is strong
evidence that she also would also accept the bet in the black room.  After
all, the situation (including the bets) is entirely symmetric between white
and black, so it is hard to see why Beauty would accept the bet in the
white room but not in the black room.  Similarly, not accepting the bet is
strong evidence that she would also not accept it in the black room.  Now,
her credence is $2/3$ that coin 2 has come up Opposite, in which case she
either will be, or has been, confronted with the black room.  Accepting the
bet now (in the white room) leads her to believe that she accepts on both
days in this case, which costs $48$ and pays off $66$, for a gain of $18$;
not accepting the bet leads her to believe that she does not accept it on
either day.  On the other hand, her credence is $1/3$ that coin 2 has come
up Grey, in which case it must now be Monday and no bet will be offered
tomorrow.  In this case, accepting the bet costs $24$ and pays off nothing,
for a loss of $24$.  Thus, in expectation, the gain from accepting the bet
is $(2/3) \cdot 18 - (1/3) \cdot 24 = 12 - 8 = 4 > 0$.  So she will accept
the bet in the white room!  Of course, by the symmetry between white and
black, this means that she will also accept the bet in the black room.
Hence, the evidential decision theorist falls for the Dutch book.

Some intuition for what makes this Dutch book work is as follows.  From the
perspective of maximizing expected net gain, clearly bet 1 is a good one to
accept, resulting in an ex ante expected net gain of
$(2/4) \cdot 22 - (2/4) \cdot 20 = 1$.  This suggests that the evidential
decision theorist's mistake is in accepting bet 2.  Always accepting bet 2
results in an ex ante expected net loss of
$(2/4) \cdot 24 - (2/4) \cdot 18 = 3$.  So what makes the evidential
decision theorist accept this bet?  Suppose she is in a white room.  She
will reason that if she accepts, then she would also accept in a black
room.  There are three possible worlds where she is in a black room at some
point: WO, BO, and BG.  But BG is ruled out by the evidence of currently
being in a white room and therefore does not factor into her current
expected payoff calculation.  Moreover, this is precisely the one world
where accepting the bet in a black room comes at a cost!  Therefore, she
evaluates the quality of the bet based on a biased selection of the
centered worlds in a black room, making the bet look better than it is.
The causal decision theorist, on the other hand, ignores bets in black
rooms altogether when making a decision in a white room, and thereby avoids
being affected by this selection bias.

\section{Discussion}

What has gone wrong for the evidential decision theorist, particularly the
evidential decision theorist who is a halfer and is hence supposed to be
immune to Dutch books according to~\cite{Briggs10:Putting}? In this
section, I first discuss the key technical problem with attempting to apply
Briggs' proof in the context of the WBG variant.  As noted earlier, a key
issue is the possibility that knowledge of a decision in one information
state affects beliefs about decisions in a different information state.
One way around the Dutch book, therefore, is to deny the possibility of
beliefs being affected in this way. I continue by arguing that this escape
route is unreasonable.  I conclude this section by discussing to what
extent susceptibility to the Dutch book indicates irrationality.

\subsection{The problem with attempting to apply Briggs' proof}

Why does Briggs' proof of the invulnerability to Dutch books of evidential
decision theorists who are halfers not apply here?  To appreciate this, it
will be helpful to first discuss some essential features of her proof.  As
noted earlier, it implicitly assumes that the information that Beauty has
available to her upon awakening during the experiment is always the same.
Briggs uses $N_W$ to refer to the number of centers (awakening events
within the experiment) in possible (uncentered) world $W$.  Suppose Beauty
is considering a bet whose net payout (including the initial cost of the
bet) is $X_W$ in world $W$.  If she is an evidential decision theorist, she
will reason that if she accepts (rejects) the bet now, then she also
accepts (rejects) it on all other occasions.  She concludes that her net
payout is $N_W X_W$ if she accepts, and $0$ if she rejects.\footnote{This
  assumes that she will be offered the same bet upon each awakening, but
  this is a reasonable requirement: see Subsection~\ref{su:legitimate}.}
Of course she does not necessarily know in which possible world she is, so
she has to consider the expected value.  Letting $\mathit{Cr}_u$ denote
halfer credences, an evidential decision theorist who is a halfer will
accept the bet if $\sum_W \mathit{Cr}_u(W) N_W X_W > 0$.

In contrast, a causal decision theorist deciding on the same bet will not
let the other $N_W-1$ bets that she is offered in world $W$ affect her
decision, so that $N_W X_W$ is replaced by $X_W$ in the above.  However, if
she is a thirder rather than a halfer, then her credence in world $W$ will
be proportional to $\mathit{Cr}_u(W) N_W$ rather than $\mathit{Cr}_u(W)$.
Hence, again, she accepts the bet if $\sum_W \mathit{Cr}_u(W) N_W X_W > 0$,
the only difference being that the factor $N_W$ comes from the credence in
this case.  Briggs proves that betting in this way (``betting at thirder
odds'') leaves Beauty immune to Dutch books.

All of this makes sense when Beauty always has the same information upon
awakening.  When this is not the case, we should first enrich the notation
a bit.  What is relevant is not the total number of centers $N_W$ in a
world, but rather the number of centers $N_W^I$ consistent with the current
information $I$.  For example, in the WBG variant, it does not suffice to
know that $N_{\text{WG}} = 2$; rather, we need that
$N_{\text{WG}}^\text{white} = 1$, $N_{\text{WG}}^\text{grey} = 1$, and
$N_{\text{WG}}^\text{black} = 0$.  Then, one might suppose that with
information $I$, the credence in some world $W$ (that is not yet ruled out
by $I$) is $\mathit{Cr}_u(W)$ for the halfer and proportional to
$\mathit{Cr}_u(W) N_W^I$ for the thirder.  (Note that this would be
consistent with the credences in the WBG variant.)  The causal decision
theorist who is a thirder would then accept the bet if
$\sum_W \mathit{Cr}_u(W) N_W^I X_W > 0$.  Now, what about the evidential
decision theorist who is a halfer?  {\em Suppose} it were the case that now
accepting (rejecting) the bet with information $I$ leads her to believe
that she always accepts (rejects) it with information $I$, {\em but does
  not influence her beliefs about what she would do given any other
  information}.\footnote{Again, note that she should always be offered the
  same bet whenever she has information $I$; otherwise, the bet offered
  would in fact give her additional information. See
  Subsection~\ref{su:legitimate} for further discussion.}  Then, in world
$W$, she believes her net payout if she rejects the bet is $c$ (i.e.,
whatever she expects to get from any bets accepted when she has information
other than $I$), and her net payout if she accepts the bet is
$N_W^I X_W + c$.  The $c$ term cancels out, and hence, again, she will
accept the bet if $\sum_W \mathit{Cr}_u(W) N_W^I X_W > 0$.  So the argument
would appear to carry through.  The problem is that, as I have argued (and
will argue further in Subsection~\ref{su:otherinfostates}), it is
unreasonable to suppose that the decision made with the current information
does not affect beliefs about decisions made with slightly different
information!  If it does affect them, then the equivalence argument falls
apart: we can no longer cancel out the $c$ term in the above because it now
depends on the decision made with information $I$, and as a result the
condition for accepting a bet changes in the case of the evidential
decision theorist who is a halfer.  This is what allows the Dutch book for
the WBG variant.  (It is worth emphasizing again that in the WBG variant,
the credences of $1/3$ placed in each of the remaining possible worlds do
not seem in question, suggesting that the problem in fact lies with
evidential decision theory, not with halfing -- at least as far as this
particular Dutch book is concerned.)

Thus, a causal decision theorist who is a thirder might analyze the
evidential decision theorist's susceptibility to the Dutch book as follows.
When an evidential decision theorist considers the payoff she expects to
get from a given action in some centered world $C$, what happens is that,
for each possible world $W$ with multiple centers that are {\em like} $C$,
she counts the effects of the action multiple times in $W$.  This is a
mistake.  On the other hand, when a halfer in centered world $C$ assesses
the probability of a possible world $W$ that contains multiple centers that
are {\em indistinguishable} from $C$, she fails to give probability to $W$
that is proportional to the number of such centers in $W$.  This, too, is a
mistake.  However, the two mistakes happen to cancel each other out
exactly, {\em if} the only centers that are {\em like} $C$ are the centers
that are {\em indistinguishable} from $C$.  Typical Sleeping Beauty
variants have this feature, giving rise to the idea that halfers who adopt
evidential decision theory avoid Dutch books.  However, it is possible for
two centered worlds to be {\em alike} while simultaneously being {\em
  distinguishable}.  This is what is happening in the WBG variant -- white
centered worlds and black centered worlds are alike but distinguishable.
Because of this, the mistake in assessing actions' payoffs is still made,
but it is not canceled out because no mistake is made in assessing the
probabilities of possible worlds.

\subsection{Avoiding the Dutch book by not changing beliefs in other
  information states}
\label{su:otherinfostates}

Based on the above, one strategy for the evidential decision theorist to
avoid the Dutch book is to never let decisions in one information state
affect beliefs about decisions in different information states.  Suppose
she takes this approach and we vary the decision that she makes in the
current centered world.  As we do so, her beliefs (conditional on this
decision) about what she would do in centered worlds that are
indistinguishable from the current one will also vary, but her beliefs
about what she would do in other centered worlds (in particular, ones that
are alike but distinguishable) will not.  This indeed avoids the Dutch
book.  But this approach seems highly unappealing.

Of course, one could add details to the case to make this approach seem
more palatable.  For example, we may suppose that before the experiment, a
neurological examination revealed to Beauty that the part of her brain that
is activated to make decisions in white rooms is entirely disjoint from the
part activated to make decisions in black rooms.  With these (or perhaps
alternative) additional details, it can perhaps be successfully argued that
her beliefs about what she does in black rooms should not be affected by
information about what she does in white rooms.  But this fails to get the
evidential decision theorist out of trouble.  The Dutch book does not need
to succeed no matter what details are added to the case.  For it to exhibit
a problem with evidential decision theory, all that is necessary is that is
succeeds for {\em some} details.  We may just as well specify that the
neurological examination reveals to Beauty that the part of her brain
involved in decision making is entirely uninfluenced by the color of the
room.  In that case, it seems entirely unreasonable for beliefs about
black-room decisions to be uninfluenced by knowledge of white-room
decisions.

Moreover, even though this is in fact not necessary for the argument to
succeed, I would argue that when {\em no} details are added to the case --
i.e., Beauty does not have any additional relevant information, such as the
results of a neurological examination -- by default, beliefs about
black-room decisions should be influenced by knowledge of white-room
decisions as I have suggested.  By way of analogy, suppose we see Kim
treating another person kindly, and this other person happens to stand to
her left.  Clearly, this will increase our credence that Kim would treat
other people who stand to her left kindly.  But it would be preposterous to
not also increase our credence that Kim would treat people who stand to her
right kindly, unless we have reason to believe that there is a fundamental
asymmetry between left and right (e.g., if we know Kim does not hear well
with her right ear and this causes her great frustration).  The situation
is similar in our context: unless we have a particular reason to believe
that the color of the room is relevant to the decisions (as in the first
example of a neurological examination), the Dutch book goes
through.\footnote{One might also suppose that knowledge of her decision in
  a white room makes Beauty only (say) 99\% confident in what her decision
  would be in a black room, where the remaining 1\% is intended to capture
  a small probability that the room color is somehow relevant to the
  decision.  It is easy to see that the Dutch book still goes through under
  these conditions.}

\subsection{Why susceptibility to (certain) Dutch books poses a problem}
\label{su:legitimate}

Does susceptibility to diachronic Dutch books really indicate
irrationality?  This question has been discussed at length in earlier
work~\citep{Hitchcock04:Beauty,Briggs10:Putting}, and I do not have much to
add that is new, but it is worth revisiting the key points here.  Some
Dutch book arguments have been made that require the bookie to have
information that Beauty does not.  For example, consider the following
Dutch book argument against a thirder, presented
by~\cite{Hitchcock04:Beauty} precisely in order to highlight this issue.
On Sunday, Beauty is offered a bet that pays out 30 on Heads, which costs
15.  Then, on Monday, Beauty is offered a bet that pays out 30 on Tails,
which costs 20.  The argument is that she is willing to accept both bets --
in particular, she is willing to accept the latter because she at that
point believes the probability of Tails is $2/3$ -- but accepting both bets
is sure to result in a loss of 5. The problem is that offering the second
bet requires the bookie to know something that Beauty does not, namely,
that it is Monday and not Tuesday.  If he did not know this, he could end
up selling the second bet to her twice in the Tails world, which would
cause the Dutch book to fail.\footnote{As already pointed out by Hitchcock,
  to be precise, what information the bookie has is not exactly what is at
  issue.  If the bookie does not know what day it is, but someone else
  prevents the bookie from offering the second bet on Tuesday to make the
  Dutch book work, this is just as problematic.  The point is that the {\em
    process as a whole} by which bets are offered to Beauty cannot use
  information that is unavailable to Beauty.}  Now, susceptibility to being
Dutch-booked by a bookie that has additional information does not seem to
indicate a failure of rationality.  After all, consider the extreme case
where a deceptive bookie already knows the outcomes of the bets in advance.
In such a case it is not surprising that he can choose which bet to offer
to ensure himself a profit.  (For a concrete example,
see~\cite{Briggs10:Putting}.)  Another take on this is that if Beauty is
astute, then the event of being offered a bet in fact provides her with
additional information, which she should take into account when deciding
whether to accept.  If our thirder Beauty above knows that the second bet
is only offered to her on Mondays, then upon being offered the bet she will
know it is Monday, update her credence in Tails to $1/2$, and not take the
bet.

However, like Hitchcock's Dutch book, the Dutch book presented in this
paper does not require the bookie to have superior information.  It is
sufficient for the bookie to know what Beauty knows (i.e., the color of the
room) in order to decide which bet to offer her.  In Hitchcock's words, he
can ``sleep with her'' -- that is, be put to sleep and awoken and have his
memory impaired in exactly the same manner.  As a result, being offered a
bet can never provide Beauty with additional information.  This remains
true {\em even if she is told the bookie's entire betting strategy at the
  outset}.  This is perhaps what most strongly suggests that susceptibility
to such a Dutch book indicates a degree of irrationality: even if Beauty is
completely aware of the game the bookie is playing with her, she still
falls for the sure loss.

\section{Concluding Remarks}

The evidential decision theorist may hope that the type of Dutch book
presented here is inherently restricted to scenarios where the agent's
memory is impaired.  But I believe that the problem runs at least a bit
deeper than that.  For example, we can easily modify the WBG variant so
that there are now two Beauties, one (``White'') who is awoken whenever the
original Beauty was awoken in the white room, and one (``Black'') who is
awoken whenever the original Beauty was awoken in the black room.  These
Beauties can be awoken simultaneously (in separate rooms) rather than
sequentially, thereby combining the WO and BO worlds into a single WB
world.  The table in Figure~\ref{fi:twobeauties} summarizes this variant.
\begin{figure}
\begin{center}
\begin{tabular}{|r|c|c|c|}\hline
& W ($1/4$) & WB ($1/2$) & B ($1/4$) \\ \hline
Monday & White awakens & both awaken & Black awakens\\ \hline
\end{tabular}
\end{center}
\caption{Possible worlds and their probabilities with two Beauties.}
\label{fi:twobeauties}
\end{figure}
We can then let the two Beauties bet under a joint account whose value they
are both trying to maximize, and, if they are evidential decision
theorists, they will fall prey to the same Dutch book, even without memory
impairment (assuming no communication between them).  The table in
Figure~\ref{fi:twobeautiesbook} summarizes the Dutch book.  For the purpose
of symmetry, we split bet 1 into two halves, each denoted 1', one for White
and one for Black, with half the cost and half the payout each.
\begin{figure}
\begin{center}
\begin{tabular}{|r|c|c|c|}\hline
  & W ($1/4$) & WB ($1/2$) & B ($1/4$) \\ \hline
  Sunday & bet 1' (W): 11 & bet 1' (W): -10 & bet 1' (W): 11\\ 
  & bet 1' (B): 11 & bet 1' (B): -10 & bet 1' (B): 11\\ \hline
  Monday & bet 2 (W): -24 & bet 2 (W): 9 & (no bet for W)\\ 
  & (no bet for B) & bet 2 (B): 9 & bet 2 (B): -24\\ \hline
  total gain from accepting all bets & -2 & -2 & -2\\ \hline
\end{tabular}
\end{center}
\caption{The table shows which bet is offered when and (in parentheses) to
  whom, as well as the net gain from accepting the bet in the corresponding
  possible world, for the Dutch book adapted to the two-Beauties variant.}
\label{fi:twobeautiesbook}
\end{figure}
Note that in this context, my interpretation of evidential decision theory
comes down to requiring that each Beauty, when calculating her expected
utility from accepting a bet (or declining it), assumes in this calculation
that the other Beauty would do the same -- so we should assume that the
Beauties are psychological twins.\footnote{Variants of the Sleeping Beauty
  problem with clones are fairly common -- see,
  e.g.,~\cite{Elga04:Defeating} and~\cite{Schwarz14:Belief}.}
Specifically, when offered bet 2, White places credence $2/3$ in the world
WB, and when calculating the value of accepting bet 2 in this world assumes
that Black will accept too, for a total gain of $2 \cdot 9 = 18$; she
places credence $1/3$ in the world W, where Black will not get to act, so
that the loss from accepting the bet is only $24$.  Because (once again)
$(2/3) \cdot 18 - (1/3) \cdot 24 = 12 - 8 = 4 > 0$, she accepts the bet
(and Black will do so as well, by symmetry).

Does the fact that the two Beauties together are susceptible to a Dutch
book indicate that they are irrational?  This is certainly not as well
established as it is for the case of a Dutch book for a single agent (as
summarized in Subsection~\ref{su:legitimate}), and some skepticism is in
order.  For example, it is well known in game theory that rational behavior
by multiple agents can result in an outcome that is strongly Pareto
dominated, i.e., there exists another outcome that all agents would
strictly prefer.  The Prisoner's Dilemma is the standard example.  However,
such examples rely on the agents having different preferences.  In
contrast, the two Beauties above have the exact same preferences.  Also, it
seems that the key properties that make a Dutch book convincing, as
discussed in Subsection~\ref{su:legitimate}, still hold here.  It is true
that the bookie will have more information than either single Beauty alone.
However, this is easily fixed by stipulating that there are {\em two}
bookies, also with a joint account, each of whom is assigned to sleep with
and offer bets to one of the Beauties.  Then, again, being offered a bet
does not provide either Beauty with more information, and this remains true
even if the bookies' joint betting strategy is common knowledge at the
outset.

It appears, then, that the Dutch book argument presented in this paper
deals a serious blow to evidential decision theory.  Of course, evidential
decision theory is often applied in settings where a decision provides
evidence not about past or future decisions that are similar, or about
decisions by another similar agent, but rather about something relevant in
the environment -- various types of brain lesion, a demon's decision, etc.
It does not appear that this Dutch book argument can be applied in such
cases, so perhaps the evidential decision theorist can retreat to an
appropriately restricted version of the theory (though it is not
immediately clear whether and how this can be coherently done).  Even upon
such retreat, I believe the glancing blow to what remains of the theory
should be cause for concern for anyone sympathetic to it.

\section*{Acknowledgments}

I thank the anonymous reviewers for many useful comments that have helped
to significantly improve the paper.

\bibliography{beauty}
\bibliographystyle{plainnat}

\end{document}